# Proton Fraction in PCR Mass Composition at Energies of $10^{15}$ - $10^{17}$ eV (Experiment «Pamir»)

Z.M. Guseva[1], E.A. Kanevskaya[1], J. Kempa[2], V.M. Maximenko[1], R.A. Mukhamedshin[3], S.E. Pyatovsky[1], V.S. Puchkov[1], S.A. Slavatinsky[1], and others

[1] *P.N. Lebedev Physical Institute of RAS, Moscow, Russia*
[2] *Warsaw Polytechnical Institute, Dep. of Mat. and Phys., Plock, Poland*
[3] *Institute for Nuclear Research, Moscow, Russia*

**Abstract**

The fraction of halo events among gamma-hadron families with $\Sigma E_\gamma > 500$ TeV is analyzed. The comparison of experimental data and Monte-Carlo simulations by MC0 code of quark-gluon string model under various assumptions on the PCR mass composition suggests a slowly increasing contribution of nuclei heavier than protons and $\alpha$-particles to PCR mass composition. Effective atomic number of PCR particles changes in the energy interval $10^{15}$ - $10^{17}$ eV so that lnA grows from 2 to 2.5.

## 1. Introduction

This paper is devoted to the investigation of gamma-families with total energy $\Sigma E_\gamma \geq 500$ TeV («super-families»), recorded in X-ray emulsion chambers (XREC) of Experiment «Pamir» at an altitude of 4370 m a.s.l. (600 g/cm$^2$) [1]. Experimental data are compared with model sampling by code MC0 of the quark-gluon model [2] which satisfactorily describes the main features of gamma-families with $\Sigma E_\gamma = 100$ - 400 TeV (primary energies $\leq 10$ PeV) under assumption of normal PCR mass composition (fraction of protons and $\alpha$-particles at an energy $E_0 = 1$ PeV constitutes 53 %). With increasing of PCR energy and $\Sigma E_\gamma$, the cascades induced by high-energy gamma-rays begin to overlap each other and, as a result, in the central part of super-families, recorded in the X-ray chamber, a diffuse region of high optical density, (so called halo) arise, its area $S$ amounts up to several centimeters squared. The calculations suggest that halo is induced, as a rule, by a narrow beam of high-energy particles, which incident upon the XREC from the atmosphere and provides the necessary density of energy flow ($\geq 20$ TeV $mm^{-2}$). The beam may be produced by development both nuclear or electromagnetic cascade in the atmosphere. Thus a halo area $S$ reflects the density of energy flow in the central part of EAS core. Fraction of halo events increases with growth of family energy and at $\Sigma E_\gamma \geq 1000$ TeV amounts up to 70 %.

Comprehensive calculations of super-families ($\Sigma E_\gamma \geq 500$ TeV) by MC0 code show that ~ 75 % of all events are produced by primary protons while in the case of halo events their fraction amounts up to 85 %. Thus the analysis of super-families provides a good approach to estimation of proton fraction in the PCR mass composition at energy $\geq 10$ PeV. Previously we show [3] that MC0 code satisfactorily describes the total intensity and halo area spectrum in gamma-families in the range of PCR energies $5 - 10^2$ PeV, provided that the PCR intensity $N(E_0 \geq 10\ PeV) = 3\times10^{-8}\ m^{-2}\ sec^{-1}\ ster^{-1}$. As this takes place the proton fraction in the PCR composition at an energy $\geq 10$ PeV is $(25 \pm 4)$ %.

In this paper we analyze a dependences of halo-event fraction on $\Sigma E_\gamma$ and halo lateral structure which are independent of PCR total intensity and are sensitive to the nature of PCR particle.

## 2. Experimental data

We analyze 143 super-families (including 60 halo events) recorded in Experiment «Pamir» during exposure $\sim 3000\ m^2\ year$ in a thin XRECs (6 cm of lead) which were exposed both as a separate unit or upper part of Pb-C XREC. Densitograms of halo were measured on X-ray films exposed at the depth of cascade development 5 cm of Pb (i.e. 9 - 11 c.u. with taking into account the incidence angle). We accept, as previously, the following criterion of halo existence: the area $S$ bounded by the isodence with optical density $D = 0.5$ (above the background, particle number density 0.04 $mkm^{-2}$) at the depth of 5 cm of Pb exceeds 4 $mm^2$. In the case of multi-core (structural) halo $\Sigma S_i \geq 4\ mm^2$ with $S_i \geq 1\ mm^2$.

## 3. Model calculations

Gamma-families were sampled by MC0 code with PCR energies in the range of $2\times10^{15}$ - $3\times10^{18}$ eV with the same exponent $\gamma = 2.05$ over all above PCR energy range. All particles of nuclear-electmagnetic cascade in the atmosphere were traced up threshold energy $E = 0.1$ TeV. Zenith angle at the boundary of the atmosphere varied in the interval $0 - 50°$. PCR mass composition in the MC0 model slowly enriches by heavy nuclei in the energy range $10^{15} - 10^{17}$. Fraction of protons and $\alpha$-particles decreases from 33 to 20 % and from 20 to 15 % accordingly, while fraction of Fe nuclei grows from 23 to 34 %.

## 4. Results

The calculations show that majority of halo events are produced by the PCR particles with energy more than 5 PeV (the probability family creation with halo in the energy range (2 - 5) PeV less than $10^{-3}$). At relatively small energies (around $10^{16}$ eV) the most fraction of halo events are produced by PCR protons since their penetrating ability is higher than for nuclei. Only at high PCR energy ($E_0 \approx 10^{18}$ eV) the probability of production of gamma-families with halo approaches to unity for all nuclei, and thus fraction of halo events from protons is determined by their fraction in the PCR composition, but contribution of the PCR particles with such an energy into overall flux of halo events is negligible. The calculations show also that primary protons produce one or two cores halo while nuclei create multi-core halo.

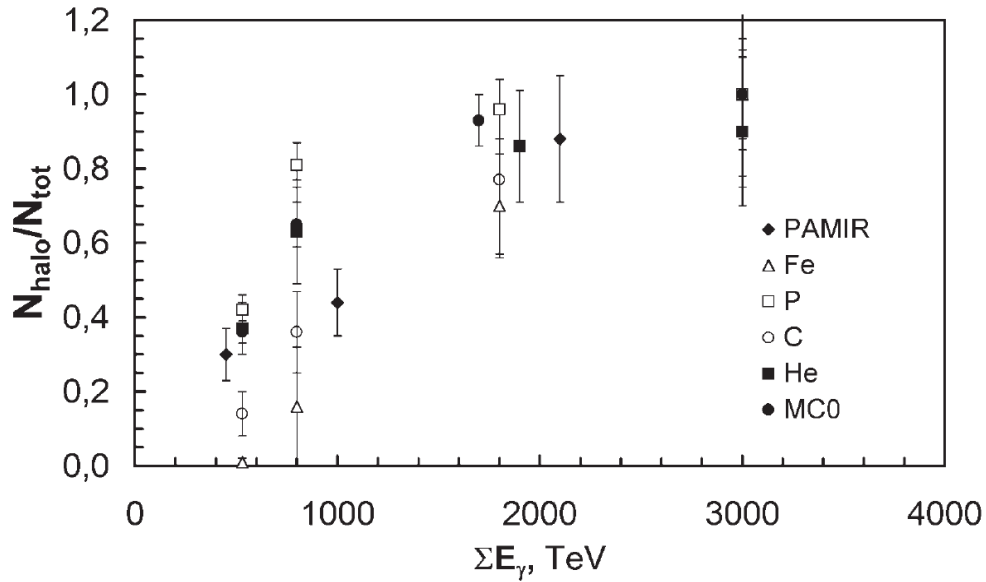

**Fig. 1.** Fraction of events with halo versus $\Sigma E_\gamma$.

Figure 1 presents the experimental and calculated (for different types of primary particles) values of fraction of events with halo $N_{halo}/N_{total}$ in dependence of $\Sigma E_\gamma$. The most interesting $\Sigma E_\gamma$ interval is 500 - 1000 TeV, where ratio $N_{halo}/N_{total}$ strongly depends from atomic number of primary particles. As is seen, experimental data are satisfactorily fitted by code MC0 and situated between calculated data, corresponding primary He and C nuclei. So average lnA at the energy $\sim 10^{16}$ eV is not more than 2.5.

In Table 1 fraction of multi-core halo calculated by MC0 code for different primary nuclei are compared with «Pamir» data. As is seen, experimental data are well fitted by MC0 model. The comparison shows that the majority of halo events are produced by primary protons.

**Table 1.** The fraction of multi-core halo in the gamma-families, produced by the different primary particles.

| p | α | C | Fe | MC0-model | Pamir |
|---|---|---|---|---|---|
| 0.25 ± 0.03 | 0.45 ± 0.09 | 0.59 ± 0.11 | 0.70 ± 0.12 | 0.28 ± 0.03 | 0.23 ± 0.07 |

**5. Conclusions**

The analysis of the super-families with $\Sigma E_\gamma \geq 500$ TeV, recorded in X-ray emulsion chambers at the Pamirs shows that:

1. Majority of the halo events are produced by primary protons with the energy $E_0 \geq 5 \times 10^{15}$ eV.
2. Dependence of the halo event fraction to all the super-families with $\Sigma E_\gamma \geq 500$ TeV from $\Sigma E_\gamma$ and fraction of the multi-core halo are satisfactorily fitted by MC0 code under the

3.     assumption that proton fraction in the PCR mass composition at the energy about $10^{16}$ eV is 25 %.

3.     It is seems that the analysis of super-families with the halo in Pamir experimental is contradict the idea about proton disappearance at the primary energy more than $10^{16}$ eV.